\documentclass[10pt]{article}

\usepackage{amsmath}
\usepackage{amssymb}

\usepackage{graphicx}

\usepackage{color} 

\usepackage{setspace} 
\usepackage[labelfont=bf,labelsep=period,justification=raggedright]{caption}

\date{}

\begin{document}

\begin{flushleft}
{\Large
\textbf{Measuring the viscous and elastic properties of single cells using video particle tracking microrheology.}
}
\\
Rebecca L. Warren, 
Manlio Tassieri$^{\ast}$, 
Xiang Li, 
Andrew Glidle, 
Allan Carlsson, 
Jonathan M. Cooper
\\
\bf{} Division of Biomedical Engineering, School of Engineering, University of Glasgow, Glasgow, U.K.
\\
$\ast$ E-mail: Manlio.Tassieri@glasgow.ac.uk
\end{flushleft}

\section*{Abstract}
We present a simple and \emph{non-invasive} experimental procedure to measure the linear viscoelastic properties of cells by passive video particle tracking microrheology. In order to do this, a generalised Langevin equation is adopted to relate the time--dependent thermal fluctuations of a bead, chemically bound to the cell's \emph{exterior}, to the frequency--dependent viscoelastic moduli of the cell. It is shown that these moduli are related to the cell's cytoskeletal structure, which in this work is changed by varying the solution osmolarity from iso-- to hypo--osmotic conditions. At high frequencies, the viscoelastic moduli frequency dependence changes from $\propto \omega^{3/4}$ found in iso--osmotic solutions to $\propto \omega^{1/2}$ in hypo--osmotic solutions; the first situation is typical of bending modes in isotropic \textit{in vitro} reconstituted F--actin networks, and the second could indicate that the restructured cytoskeleton behaves as a gel with ``\textit{dangling branches}". The insights gained from this form of rheological analysis could prove to be a valuable addition to studies that address cellular physiology and pathology.


\section*{Introduction}
As many authors have noted, the mechanical properties of a cell's cytoskeleton can influence factors such as growth, apoptosis, motility, signal transduction and gene expression \cite{Chicurel:1998xq}. Related to this, there is a desire to be able to provide a rheological interpretation of the cell's viscoelastic response that has the potential to yield quantitative information on the cell's cytoskeletal structure. Consequently, in this work, we have developed a means of using passive video particle tracking microrheology measurements to quantitatively measure changes in the viscoelastic properties of a cell as a consequence (in this case) of simple changes in its external environment, i.e. subjecting a cell to a hypo--osmotic shock.

Osmotic regulation and the transport of osmotically active molecules are fundamental to both metabolic processes and homeostasis of cells and require precise regulation and maintenance of intracellular water \cite{Nagy:2010sj}. The ability of many cells to regulate their volume, \textit{via} internal restructuring, in response to osmotic changes in their environment is an essential component of normal cellular function. This regulatory volume change is linked to a reorganisation of the cytoskeletal actin networks \cite{Papakonstanti:2007fi, Ebner:2005hq}. Exposure to a hypotonic solution typically induces an initial rapid swelling followed by a shrinkage of the cell that occurs over a slower timescale of several minutes. The modulation of the actin dynamics and polymerisation that accompanies these regulatory volume changes has been shown in various cell types by methods such as DNase I inhibition assay and fluorescence measurements of phalloidin--labelled actin \cite{Papakonstanti:2007fi}.

In this work we have developed a video particle--tracking tool to study the microrheology of cells, using a Jurkat lymphocyte cell line as a model system.  In general, lymphocytes have been shown to have a response to hypo--osmotic solutions that is influenced by tyrosine kinase activity \cite{Lepple-Wienhues:1998et} and cytoskeleton (F--actin) participation in ion channel activation \cite{Levitan:1995cr}. Furthermore, the interest in T--cells stems from their important role in the regulation of immune responses. In particular, Jurkat cells, as used in this study, are CD4+ T--lymphoma cells that are often utilised as models to understand T--cell signalling~\cite{Abraham:2004os} and HIV-1 dissemination in viral pathogenesis \cite{Cheng-Mayer:1989qq}.

The linear viscoelastic properties of a material can be represented by the frequency--dependent dynamic complex modulus $G^{*}(\omega)$, which provides information on both the viscous and the elastic nature of the material (i.e. on how the matter dissipates and stores the energy transferred to it) at different frequencies $\omega$; it is defined as the ratio between the Fourier transforms (denoted by the symbol `` $\hat{}$ ") of the applied stress $\sigma (t)$ (which is proportional to the applied force) and the resulting strain $\gamma (t)$ (which is proportional to the material/cell deformation):
\begin{equation}
\label{G*}
	G^*(\omega) = \frac{\hat{\sigma}(t)}{\hat{\gamma}(t)} \equiv \frac{\int_{-\infty}^{+\infty} \sigma(t)e^{-i\omega t}dt}{\int^{+\infty}_{-\infty} \gamma(t)e^{-i\omega t}dt}
\end{equation}
where $i$ is the imaginary unit (i.e. $i^2=-1$).

The standard method of measuring $G^{*}(\omega)$ is based on the imposition of an oscillatory stress $\sigma (\omega,t)$ given by a formula such as:
\begin{equation}
\label{sigma}
	\sigma (\omega,t) = \sigma_0\sin(\omega t)
\end{equation}
where $\sigma_0$ is the amplitude of the stress function, $\omega$ is the frequency and $t$ the time. The resulting oscillatory strain $\gamma (\omega,t)$ then has the formula:
\begin{equation}
\label{gamma}
	\gamma (\omega,t) = \gamma_0\sin(\omega t+\delta(\omega))
\end{equation}
where $\gamma_0$ is the strain amplitude and $\delta(\omega)$ is the frequency--dependent phase shift between the stress and the strain. The amplitudes of the complex modulus� in--phase and out--of--phase components are proportional to the ratio between amplitudes of the stress and of the strain, with constants of proportionality defining the storage (elastic) $G' (\omega)$ and the loss (viscous) $G'' (\omega)$ moduli, respectively:
\begin{equation}
\label{G*2}
	G^*(\omega) = G' (\omega)+iG'' (\omega)=\frac{\sigma_0}{\gamma_0}\cos(\delta(\omega))+i\frac{\sigma_0}{\gamma_0}\sin(\delta(\omega))
\end{equation}
For example, in the case of a purely elastic solid, the stress and the strain are in phase (i.e. Hooke's law, the material deformation is directly proportional to the applied force) and $\delta(\omega)=0$; whereas, for a purely viscous fluid, such as water or glycerol, $\delta(\omega)=\pi/2$. For complex solids (e.g. gels, cells) or viscoelastic fluids (e.g. blood, saliva) $\delta(\omega)$ would take any value between these limits (i.e. $0\leqslant\delta(\omega)\leqslant\pi/2$) depending on the frequency at which the force (stress) is applied.

\begin{figure}[!h]
\begin{center}
\includegraphics[width=5.7in]{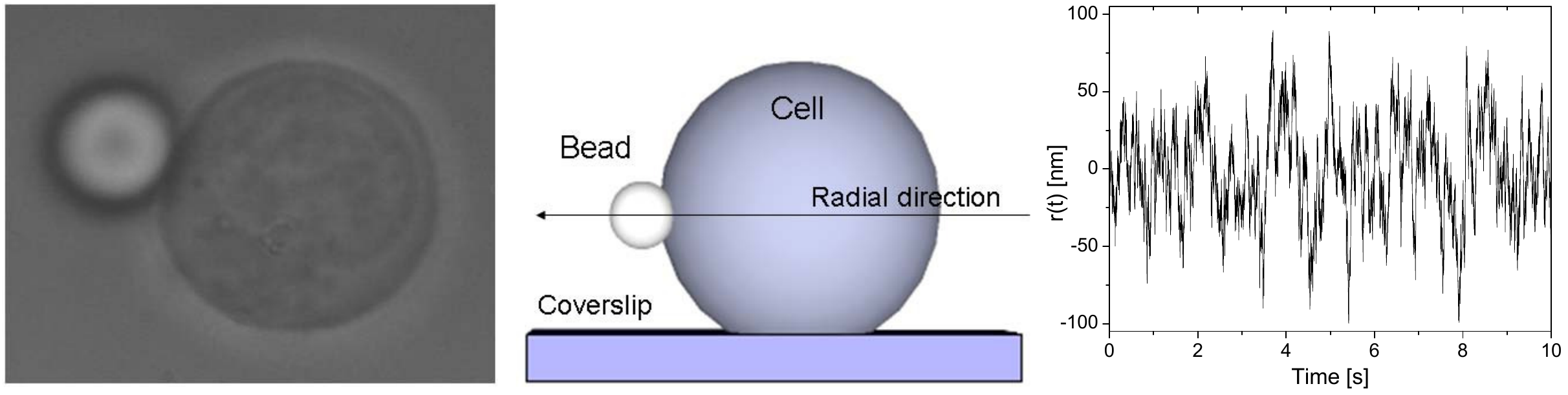}
\end{center}
\caption{
{\bf (Left) Top-view of a 5 $\mu m$ diameter silica bead bound to the surface of a Jurkat cell.}  (Centre) Schematic side-view of the left image. (Right) The thermal fluctuations of a 5 $\mu m$ silica bead chemically bound to the surface of a Jurkat cell. Images were analysed in real--time and the coordinates of the bead's centre of mass were stored directly on the random--access memory (RAM) of the computer, at time intervals down to $1/600 s$, using our own suite of image analysis software written in LabVIEW~\cite{Gibson:2008zh}.}
\label{Figure_1}
\end{figure}

The aim of this article is to present a straightforward and \emph{non--invasive} procedure for measuring the \textit{in vivo} linear viscoelastic properties of single cells via passive video particle tracking microrheology of single beads attached to the cells' exterior. This method has advantages over both complicated active microrheology techniques, where complex experimental set--ups are necessary to exert an external force for performing stress--controlled measurements; and invasive passive video particle tracking of submicron--probes embedded (either \textit{via} endocytosis or micropipette injection) within the cell's cytoskeleton (e.g. \cite{Bausch:1998ee, Bausch:2001xw, Deng:2006pt, Hoh:1994ta, Uhde:2005pz, Wirtz:2009sh, Lu:2006jh}. In particular, the procedure consists of measuring the thermal fluctuations of a bead chemically bound to the cell's exterior (Figure~\ref{Figure_1}), for a sufficiently long time. A generalised Langevin equation was adopted to relate the time--dependent bead trajectory, $\vec{r}(t)$, to the frequency--dependent moduli of the cell. Notably, the procedure presented here represents an alternative methodology that can be extended to many experimental formats and provides a simple addition to existing cellular physiology studies (e.g. those monitoring cell pharmacological response). Indeed, when compared to single cell viscoelasticity assays such as magnetic tweezers, atomic force microscopy and optical stretcher, as employed in \cite{Bausch:1998ee, Bausch:2001xw, Deng:2006pt, Hoh:1994ta, Uhde:2005pz, Wirtz:2009sh, Lu:2006jh}, our method has the advantage of revealing the changes of the cell's viscoelastic properties over a wide range of frequencies (here from $\thicksim$0.6 Hz up to $\thicksim$600 Hz), to a high level of accuracy, whilst it experiences an induced physiological process.

\section*{Analytical model}
The equation describing the \textit{pseudo} Brownian fluctuation of the randomly varying bead position $\vec{r}(t) \forall t$ can be derived by means of a generalised Langevin equation, which was originally derived to describe the Brownian motion of a freely diffusing particle suspended in a purely viscous fluid \cite{Langevin:1908fk} and represents Newton's second law as written for stochastic processes:
\begin{equation}
\label{Langevin}
m\vec{a}(t)=\vec{f}_{R}(t)-\int^{t}_{0}[\zeta_{c}(t-\tau)+\zeta_{s}(t-\tau)]{\vec{v}} (\tau)d\tau
\end{equation}
where $m$ is the mass of the particle, $\vec{a}(t)$ is its acceleration, $\vec{v}(t)$ is the bead velocity and $\vec{f}_{R}(t)$ is the usual Gaussian white noise term that models stochastic thermal forces acting on the particle. The integral term represents the total damping force acting on the bead, which, based on the superposition principle, incorporates two generalised time--dependent memory functions $\zeta_{c}(t)$ and $\zeta_{s}(t)$ that are representative of the viscoelastic nature of the cell and the solvent, respectively. These memory functions are directly related to the materials' complex modulus as shown hereafter.

In the case of the solvent, which in this work is the media surrounding the cell, the relationship between memory function and complex modulus is straightforward. Indeed, to a first approximation, using the assumptions adopted by Mason and Weitz \cite{Mason:1995px} when studying the motion of thermally excited free particles, at thermal equilibrium the complex viscoelastic modulus of the solvent is related to the memory function $\zeta_{s}(t)$ by the expression: $G^*_s (\omega)=i\omega \hat{\zeta}_{s}(\omega)/6\pi R$, where $R$ is the bead radius and $\hat{\zeta}_{s}(\omega)$ is the Fourier transform of $\zeta_{s}(t)$. In the case of the cell, we assume a similar relationship between $G^*_c (\omega)$ and $\zeta_{c}(t)$ to that given above for the solvent, but with a different constant of proportionality that we will call $\beta$; i.e., $G^*_c (\omega)=i\omega \hat{\zeta}_{c}(\omega)/\beta$. Note that, $\beta$ may vary for different cells because it depends on ($i$) the cell radius ($R_c$), ($ii$) the number of the chemical bonds between the bead and the cell, ($iii$) the contact area between the cell and the glass coverslip, and ($iv$) the relative position of the bead with respect to both the cell's equatorial plane and the glass coverslip. In this study we will only focus on the changes in cell dynamics due to variations in osmolarity. This can be achieved since all the above experimental parameters, with the exception of $R_c$, are (in good approximation) time--independent, and will not change significantly during the course of a set of measurements, as shown in Figure~\ref{Figure_4} for measurements performed on cells in iso--osmotic condition. In addition, it is important to highlight that the cells maintained a spherical shape well beyond the duration of the experiment. Moreover, the cells adhered to the coverslip and no drift was observed during the experiment. Finally, given that we measure an increase of the cell radius ($R_c$) of $\leqslant 5\%$ for the hypo--osmolarity experiments described below, we also make the assumption that this small change does not affect the dynamics of the system appreciably.

We now describe how the thermal fluctuations of a bead, chemically bound to a cell, can be investigated to determine the viscoelastic properties of the cell through the analysis of the time dependence of the bead's mean--square displacement (MSD) $\left\langle \Delta r^{2}(\tau)\right\rangle \equiv \left\langle\left[\vec{r}(t+\tau)-\vec{r}(t)\right]^{2}\right\rangle_{t},$ in the radial direction of the cell; where $t$ is the absolute time and $\tau$ is the lag--time (i.e. time interval). The average is taken over all initial times $t$. Under these circumstances, it is an easy step to show that, at thermal equilibrium, Equation~\ref{Langevin} can be reorganised to express the viscoelastic moduli of the cell as function of the Fourier transform of the mean--square displacement:
\begin{equation}
\label{G*c}
	\frac{G^*_c (\omega)}{G'_0} = \frac{1}{i\omega \hat{\Pi}(\omega)}+\frac{m\omega^2}{\beta G'_0}-\frac{6\pi R G^*_s (\omega)}{\beta G'_0}
\end{equation}
where $G'_0$ is the limiting value, for vanishingly low frequencies, of the elastic modulus of the whole system (i.e. cell \textit{plus} solvent), which in this work is $G'_0 \equiv G^*_c (\omega)$ for $\omega \to 0$; and $\hat{\Pi}(\omega)$ is the Fourier transform of the normalised mean--square displacement $\Pi (\tau)=\left\langle \Delta r^{2}(\tau)\right\rangle/\left\langle  r^{2}\right\rangle$, introduced by Tassieri \textit{et al.}~\cite{Tassieri:2010ri} in the study of the thermal fluctuations of an optically trapped bead suspended in a viscoelastic fluid. The term $\left\langle  r^{2}\right\rangle$ is the time--independent variance of $\vec{r}(t)$ measured for a sufficiently long time. Moreover, it has been shown~\cite{Tassieri:2010ri} that for a constrained bead (e.g. one that is optically trapped or, as in this work, chemically bonded to the cell) $\Pi (\tau)=1$ at large time intervals. It is important to highlight that (as in the case of an optically trapped bead) the term $\beta G'_0$ (representing the coefficient of the elastic restoring force) can be determined by means of the principle of equipartition of energy, which in one dimension is written as:
\begin{equation}
\label{Equipartition}
k_{B}T=\beta G'_0 \langle r^2 \rangle
\end{equation}

Usefully, two further simplifications can be made to Equation~\ref{G*c} because ($a$) for micron-sized silica beads, the inertia term $m\omega^2$ is negligible up to frequencies on the order of $MHz$ and ($b$) for solvents having frequency--independent viscosity $\eta$ (e.g. water), $G^*_s (\omega)$ simplifies to $i\omega \eta_s$. Thus, Equation~\ref{G*c} provides the viscoelastic properties of the cell (scaled by $G'_0$) over a range of frequencies that is limited at the top end by the rate of acquisition of images to determine the bead position, and at the bottom end by a cutoff frequency given by $\omega_{cutoff}=\beta G'_0/\eta_0$, where $\eta_0$ is the limiting value, for vanishing frequencies, of the viscosity of the whole system.

Finally, in order to evaluate the Fourier transform in Equation~\ref{G*c} we adopt the analytical method introduced by Evans \textit{et al.}~\cite{Evans:2009ay} (i.e. Eq. (10) of Ref.~\cite{Evans:2009ay}), which is applied directly to the experimental data points and has the advantage of removing the need for Laplace and inverse--Laplace transformations of experimental data \cite{Mason:1997bs}.

\section*{Experimental details}
Anti-CD4+ (Invitrogen) was attached to $5\mu m$ carboxylate functionalised silica beads (Bangs Laboratories Inc) using a method based on a previously described protocol \cite{Kulin:2002yb}.  The adaptation of this method involved first binding a short polyethylene glycol chain to the beads so as to provide a flexible linker on to which the anti-CD4+ could be bound.

In detail, the microspheres were washed (3x) with 0.1 M $Na_{2}CO_3$ buffer (pH 9.6) using centrifugation between washes to remove the supernatant liquid.  This ensured that all the carboxylate groups were deprotonated. Following the final wash, the beads were resuspended in RO water and centrifuged again. After removal of the water, the remaining pellet was resuspended in freshly prepared 0.02 M sodium phosphate buffer (pH 6) containing 2 mM 1-ethyl-3-(3-dimethylaminopropyl)-carbodiimide (EDC, Sigma) and 5 mM sulfo-N-hydroxysuccinimide (sulfo-NHS, Thermo-Fisher). The microsphere solution was then agitated for 15 minutes to allow the sulfo-NHS to activate the carboxylate groups on the surface of the microspheres. The solution was centrifuged, then the supernatant removed and replaced by a 10 mM solution of the flexible linker, O-(2-Aminoethyl)-O'-(2-carboxyethyl)polyethylene glycol 3,000 hydrochloride (Aldrich) in pH 7.4 phosphate buffer.  The activated microspheres were then agitated in the flexible linker solution for 2 h at 4$^oC$, before further centrifugation and washing with pH 7.4 buffer. The anti-CD4+ was then bound to the carboxylate terminals of the PEG linker from a 1 mg/ml solution of anti-CD4+, using the same EDC/sulfo-NHS activation procedure outlined above.

The Jurkat cells were obtained from ATCC (clone E6-1, TIB-152) and maintained in RPMI 1640 (1x) media containing L-Glutamine and 25mM HEPES [Gibco, Invitrogen]. Prior to the experiment they were suspended in PBS [Gibco, Invitrogen] with an osmolarity of 275-304 mOsm, then mixed with the functionalised beads and allowed to equilibrate for $\thicksim$10min.

Beads were optically trapped and moved to the equatorial plane of the cell, where they were held for a sufficient time ($\thicksim$2 min) until the binding process was completed. Trapping was achieved by using a CW Ti:sapphire laser system (M Squared, SolsTiS) which provides up to 1 W at 830 nm. Holographic optical traps were created via the use of a spatial light modulator (Boulder XY series) in the Fourier plane of the optical traps \cite{Preece:2011ct}. The microscope was equipped with a 100x 1.3NA (Zeiss, Plan-Neofluor) objective lens, which was used both to focus the trapping laser beam and to visualise the particles via a Prosilica PCI-6713 camera. Samples were mounted on a motorised microscope stage (Prior Pro-Scan II). The beads' trajectories were measured and stored on a personal computer in real-time, at frequencies up to $\thicksim$kHz (depending on the size of the analysed region of interest (ROI) within the camera field of view), using our own suite of image analysis software written in LabVIEW \cite{Gibson:2008zh}.

\section*{Results and Discussion}
In order to validate the analytical model introduced in this work, we measured the dynamic response of an analytical system that could be considered as an \emph{ideal} viscoelastic medium with known storage and loss moduli; i.e. an optically trapped bead suspended in a Newtonian fluid. Indeed, for such a system, the elastic modulus must be proportional to the optical tweezers' (OT) trap stiffness ($\kappa$), this being the only elastic component present in the system (i.e. the ensemble of Newtonian fluid, bead and OT); and the viscous modulus must be linearly proportional to the frequency ($\omega$) with constant of proportionality given by the viscosity of the Newtonian fluid ($\eta$).

\begin{figure}[!h]
\begin{center}
\includegraphics[width=5.7in]{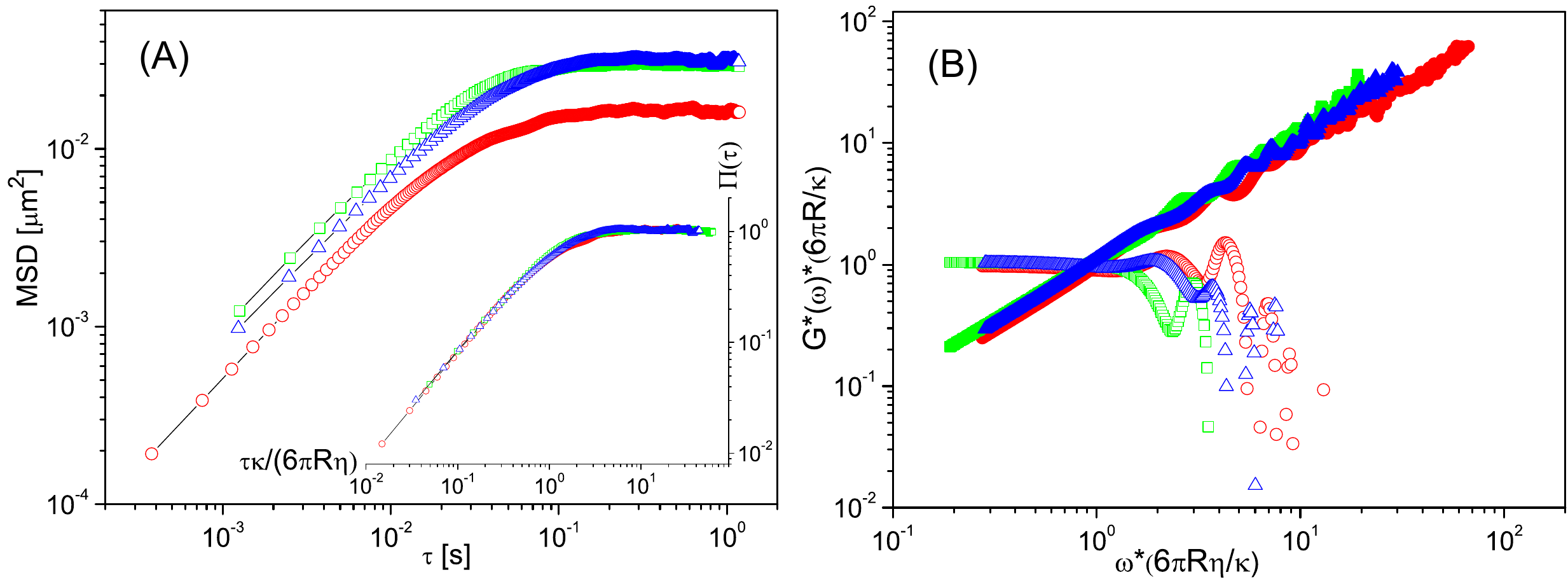}
\end{center}
\caption{
{\bf (A) The mean-square displacements \textit{vs.} lag-time of optically trapped beads in Newtonian fluids.}  The square and the triangle symbols refer to measurements performed with 2 $\mu m$ diameter silica bead in Glycerol/water mixtures at 10\% and 20\% w/w, having viscosity values of 1.15 $mPas$ and 1.54 $mPas$, and trap constants of 0.85 $\mu N/m$ and 0.81 $\mu N/m$, respectively; The circle symbols refer to a measurement performed with 5 $\mu m$ diameter silica bead in water having viscosity value of 0.89 $mPas$ and trap constants of 1.68 $\mu N/m$ (note that for this measurement only we used an off-line faster camera with internal memory; not suitable for continuous acquisition). The inset shows the same data as above but with the abscissa scaled by $\kappa/(6\pi \eta R)$. (B) The normalised viscoelastic moduli \textit{vs.} the normalised frequency for all three analytical systems described above.}
\label{Figure_2}
\end{figure}

In Figure~\ref{Figure_2}B we present the normalised viscoelastic moduli \textit{versus} the normalised frequency for three systems made up with diverse combinations of Newtonian fluids and beads having different viscosities and size, respectively. For each system, the moduli have been derived by the analysis of the thermal fluctuations of an optically trapped bead by means of Equation~\ref{G*c} \textit{via} the relative MSD shown in Figure~\ref{Figure_2}A. In this case, the term $G^*_c (\omega)$ in Equation~\ref{G*c} represents the viscoelastic nature of the analytical system; the term $G^*_s (\omega)$ simplifies to $i\omega \eta$ and is coincident with $\Im[G^*_c (\omega)]$ (and therefore it does not need to be accounted for in Eq.~\ref{G*c}); the term $\beta G'_0 \equiv \kappa$ is determined by means of the principle of equipartition of energy with $\beta \equiv 6 \pi R$. In summary, for both moduli, the normalising factor is given by $\kappa/(6\pi R)$; whereas, for the abscissa, the scaling factor is given by a cut-off frequency defined as $\kappa/(6\pi R\eta)$, which also works in the time-domain as shown in the inset of Figure~\ref{Figure_2}A. In Figure~\ref{Figure_2}B it is interesting to highlight that at low frequencies (i.e. up to the cut-off frequency) the elastic modulus $G' (\omega)$ is actually normalised to~$1$, as expected; whereas, at high frequencies (i.e. for $\omega > \kappa/(6\pi R)$) $G' (\omega)$ vanishes towards zero. The latter phenomenon is itself an interesting feature that is currently an object of discussion in literature \cite{Yanagishima:2011zh, Pesce:2009xh, Addas:2004dl}. In particular, it is argued that the analytical models adopted so far for data analysis of the thermal fluctuation of an optically trapped bead suspended into a Newtonian fluid are unable to provide the high frequency elastic component of the optical tweezers (OT), because they would ``suffer from systematic truncation errors at high frequencies". In contrast, we believe that the absence of the optical tweezers' high frequency elastic component is an \emph{intrinsic} feature of the system made up by the ensemble of (Newtonian) fluid, bead and OT~\cite{Wu:2009fk}; on which we will provide a detailed explanation, but in a different work. On the other hand, the viscous modulus $G'' (\omega)$ grows linearly with the frequency (thus in a Log-Log plot it is represented by a line with unity slope) over the entire experimental frequency window, with constant of proportionality given by the viscosity of the Newtonian fluid, as expected. Note that the results shown in Figure~\ref{Figure_2} are in good agreement with those published in literature \cite{Yanagishima:2011zh, Pesce:2009xh, Addas:2004dl} for an optically trapped bead in water, but they used a different analytical method for data analysis and they overlooked the scaling process introduced in this work.

Having demonstrated the validity of our analytical model, we now investigate its applicability to cellular physiology studies by focussing on changes in the cells' microrheology due to changes in osmolarity of the surrounding solution.

\begin{figure}[!h]
\begin{center}
\includegraphics[width=4in]{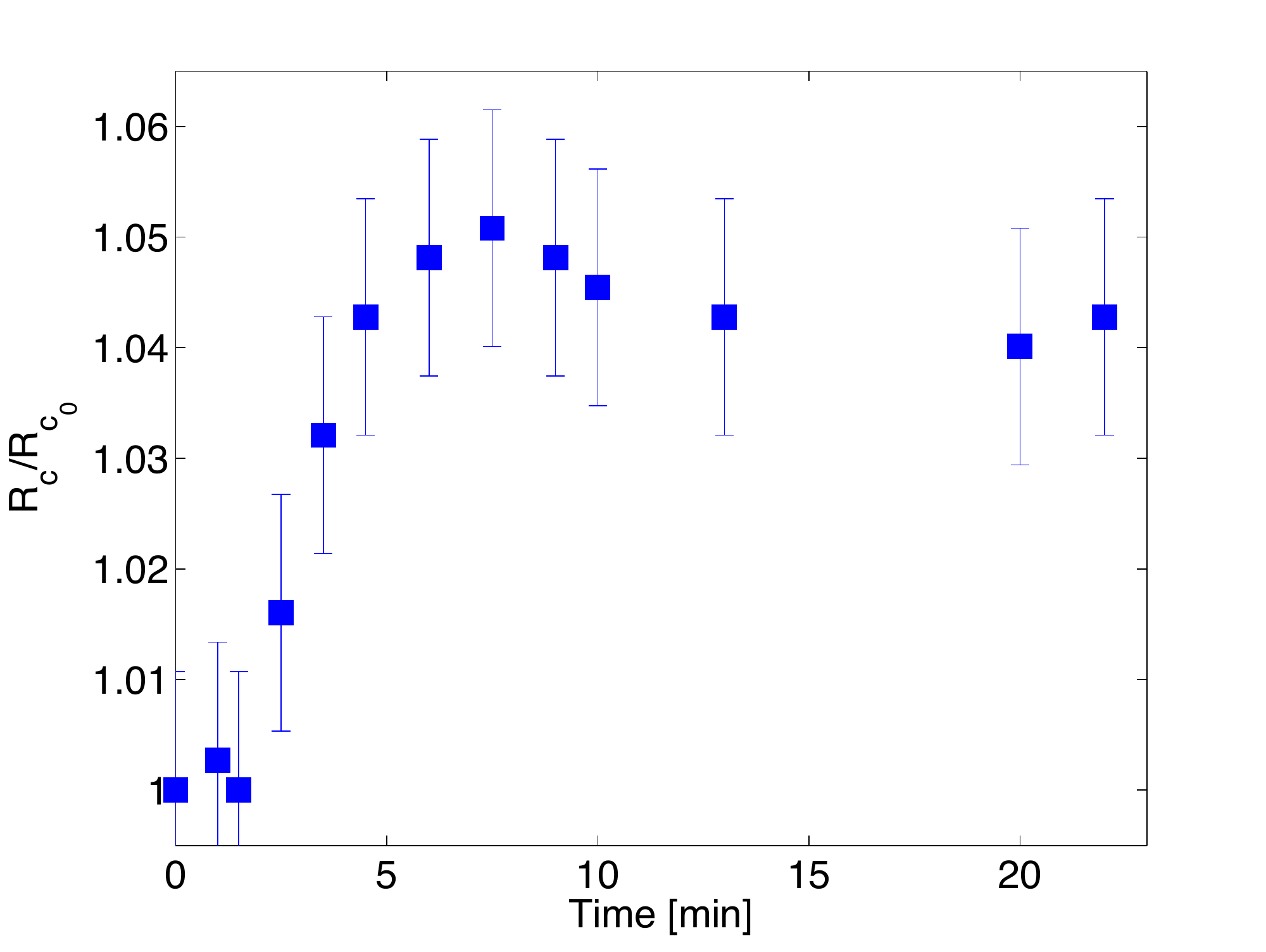}
\end{center}
\caption{
{\bf Averaged relative cell radius changes \textit{vs.} time as a consequence of the change in osmolarity of the cell solution; from iso-osmotic conditions in PBS buffer towards a hypo-osmotic condition obtained by adding 10\% v/v of distilled water to the initial solution.}  The osmotic shock causes a rapid swelling of the cells, followed by a partial recovery of the cells' volume. These results are in good agreement with those reported in literature, e.g.~\cite{Cantiello:1997kn}.}
\label{Figure_3}
\end{figure}

It is anticipated that there are two major physiological processes that accompany the changes in osmolarity of the solution surrounding the Jurkat cell: (I) an actin cytoskeleton reorganisation and (II) an increase of the cell volume as the solution becomes hypo--osmotic. In order to quantify the changes in cell volume, we assumed that the cells were spherical, Figure~\ref{Figure_1}.  In particular, Figure~\ref{Figure_3} shows that the relative radius changes are $<6\%$ when the surrounding PBS solution was made hypo-osmotic by addition of 10$\%$ v/v distilled water. The data of Figure~\ref{Figure_3} also shows that, after an initial swelling, the cells contract to a still swollen state that is $\thicksim$5$\%$ larger than when they were in isotonic PBS. Based on these observations, which are in good agreement with those already existing in literature (e.g. \cite{Cantiello:1997kn}), we assume that these small radius changes do not affect the dynamics of the system appreciably.

To study how the \textit{in vivo} viscoelastic properties of Jurkat cells vary with the osmolarity of the solution, we have adopted the model described in the Analytical Section. In particular, analysis of different time regimes in the time dependent normalised mean-squared displacement, $\Pi (\tau)$, has the potential to reveal information both on the pure elastic component of the cell (at long time intervals or low frequencies) and on the fast dynamics occurring at small length scales (e.g. those related to the transverse bending modes of single actin filaments in the cytoskeleton).

\subsection*{Elastic properties of the cell derived from $\Pi (\tau)$}
\begin{figure}[!ht]
\begin{center}
\includegraphics[width=4in]{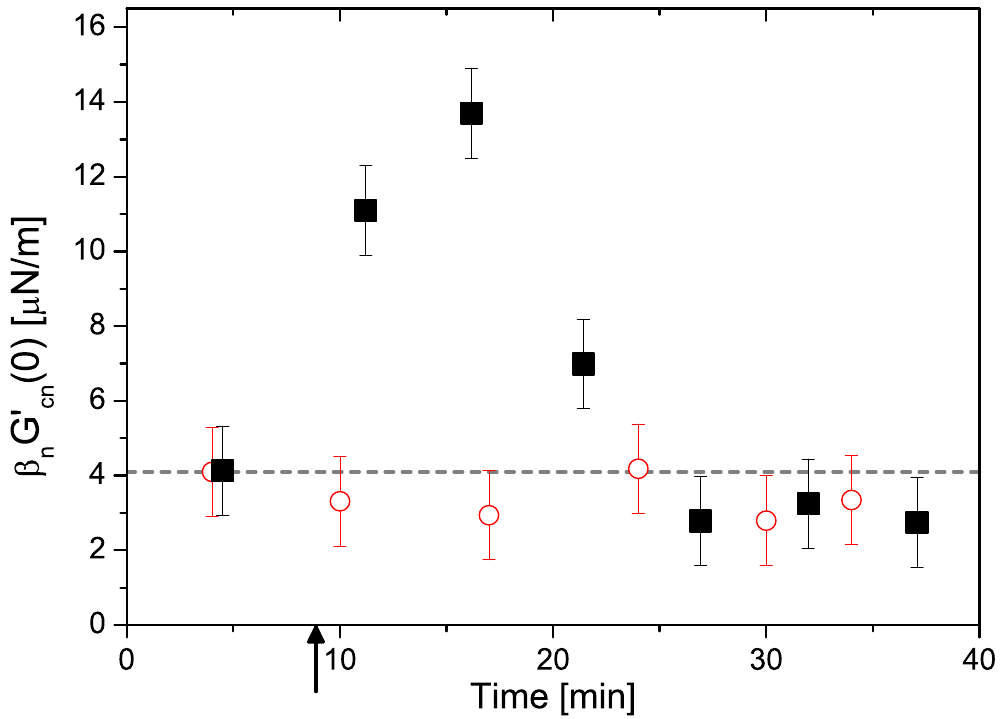}
\end{center}
\caption{
{\bf The absolute value of $\beta G'_c(0)$ measured in isotonic (PBS) condition (circle symbols) and when the solution is made 10\% hypotonic (square symbols)} The arrow indicates the time at which the isotonic solution is made hypotonic. The dotted line is a guide for the eye, so that the later points in the graph can be compared to the first points measured in isotonic condition.}
\label{Figure_4}
\end{figure}

The first result that can be obtained through the analysis of the thermal fluctuations of the bound bead (Figure~\ref{Figure_1}) is the relative change of the cell's low frequency elastic plateau modulus $G'_c (0)$, Figure~\ref{Figure_4}. This can be evaluated from the time--independent variance $\langle r^2 \rangle$ of the constrained bead in the radial direction of the cell, by adopting the principle of equipartition of energy as in Equation~\ref{Equipartition}. In particular, although the absolute value of $G'_c (0)$ can not be determined because of the unknown factor $\beta$, the relative change between different values of $\beta G'_c (0)$, obtained from $n$ sequential measurements performed on the same cell, differ from each other only because of both the small increase of the cell radius (measured as $\thicksim$5$\%$) and the variation of $G'_c (0)$. Thus, from Figure~\ref{Figure_4} it is clear that a change of 10$\%$ in osmolarity towards hypotonicity, which occurs at $t\thicksim$9min as a result of adding distilled water to the iso-osmotic solution, induces a temporary increase of $\beta G'_c (0)$ that lasts for approximately 15-30min. At its greatest, this change in the low frequency elastic modulus dependent term, $\beta G'_c (0)$, is over 3 fold in magnitude.

The explanation for such a large increase may in part be due to the change in radius; however, if this were solely responsible, it would be expected that $\beta G'_c (0)$ would remain significantly above its original value, as the cells do not regain their original radius. Thus, since $\beta G'_c (0)$ \textit{does} return to close to the original value (at $t\thicksim$25 min), consideration should be given to the temporary increase in electrostatic intermolecular interactions within the cell due to the reduction in ionic strength of the solution (i.e. a reduction in screening of the charges on cytoskeleton molecules by solution based ions). This latter phenomenon would also alter the cell's viscoelastic properties and any resulting actin cytoskeleton rearrangement may be seen as part of an attempt of the cell to re--equilibrate the solution osmolarity, as discussed below.

\begin{figure}[!ht]
\begin{center}
\includegraphics[width=4in]{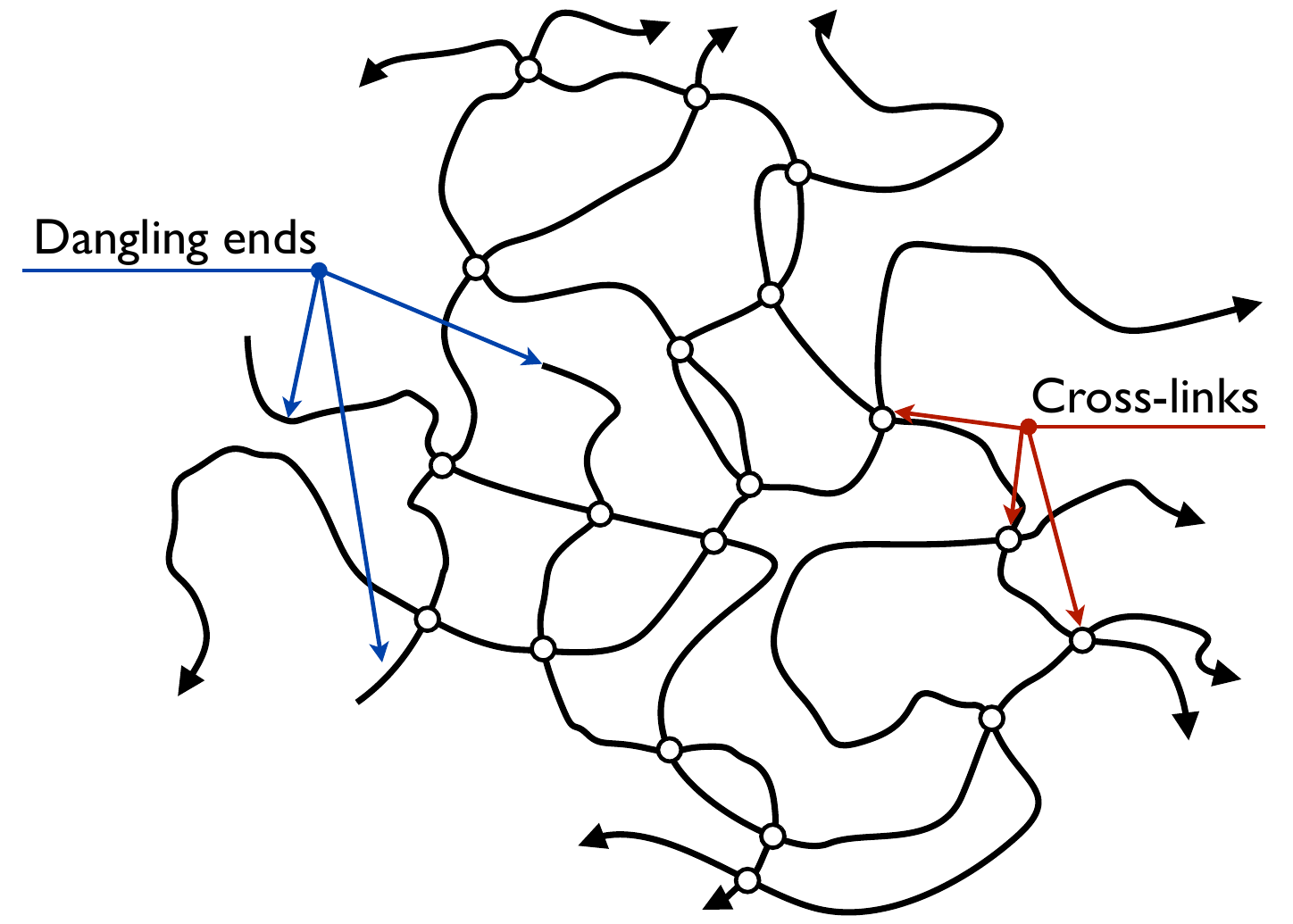}
\end{center}
\caption{
{\bf Schematic view of a randomly cross-linked network with ``\textit{dangling branches}" defects. Circles represent cross-linking junctions and arrows denote attachments to the macroscopic network.} The ``\textit{dangling branches}" do not contribute to the gel elasticity because they cannot bear stresses; in contrast, they can dissipate energy by friction with the solvent.}
\label{Figure_5}
\end{figure}

Evidence for an increase in intermolecular interactions in the cytoskeleton being the cause of the observed increase in $\beta G'_c (0)$ comes from studies of polymer gels \cite{Rouse:ya, Ferry:1980jo, Rubinstein:2003fb}, for which the elastic plateau modulus ($G' (0)$) of an ideal cross-linked polymer network can be written:
\begin{equation}
\label{Gell}
      G' (0) \cong \psi k_B T \frac{(f-2)}{f}
\end{equation}
where $k_B$ is the Boltzmann constant, $T$ is the absolute temperature, and $\psi$ and $f$ are the number of network strands and cross--links per unit volume, respectively (see Figure~\ref{Figure_5}). Here, an increase in number of strands and/or number of cross--linking interactions causes an overall increase in $G' (0)$. Note that although it is appreciated that there may be more order associated with macromolecules within the cell's cytoskeleton, the generalised polymer network of Figure~\ref{Figure_5} nevertheless has many \textit{similarities} with the complex structure of the cytoskeleton. For example, it is well known that all cells possess a host of actin--binding proteins that can align actin filaments into bundles and then cross-link filaments or bundles into networks~\cite{Pollard:1986bx}. Bundled actin and scruin \textit{in vitro} networks act similarly to cross--linked networks and show a strong increase in elasticity as the degree of bundling increases~\cite{Gardel:2004fx}.

Thus, by analogy with the models for a polymer network, a $>$3 fold increase in the cell elastic plateau modulus could be explained by an increase of intermolecular interactions (i.e. electrostatic interactions) that would have the effect of forming temporary cross--linkages and reshaping the isotropic cytoskeleton structure. This scenario is in good agreement with confocal microscopy results already reported for multiple cell lines and the observations that actin forms bundles in aortic bovine endothelial cells after hypotonic exposure~\cite{Koyama:2001zv}.

\subsection*{Fast dynamics of the cell derived from $\Pi (\tau)$}
\begin{figure}[!ht]
\begin{center}
\includegraphics[width=5in]{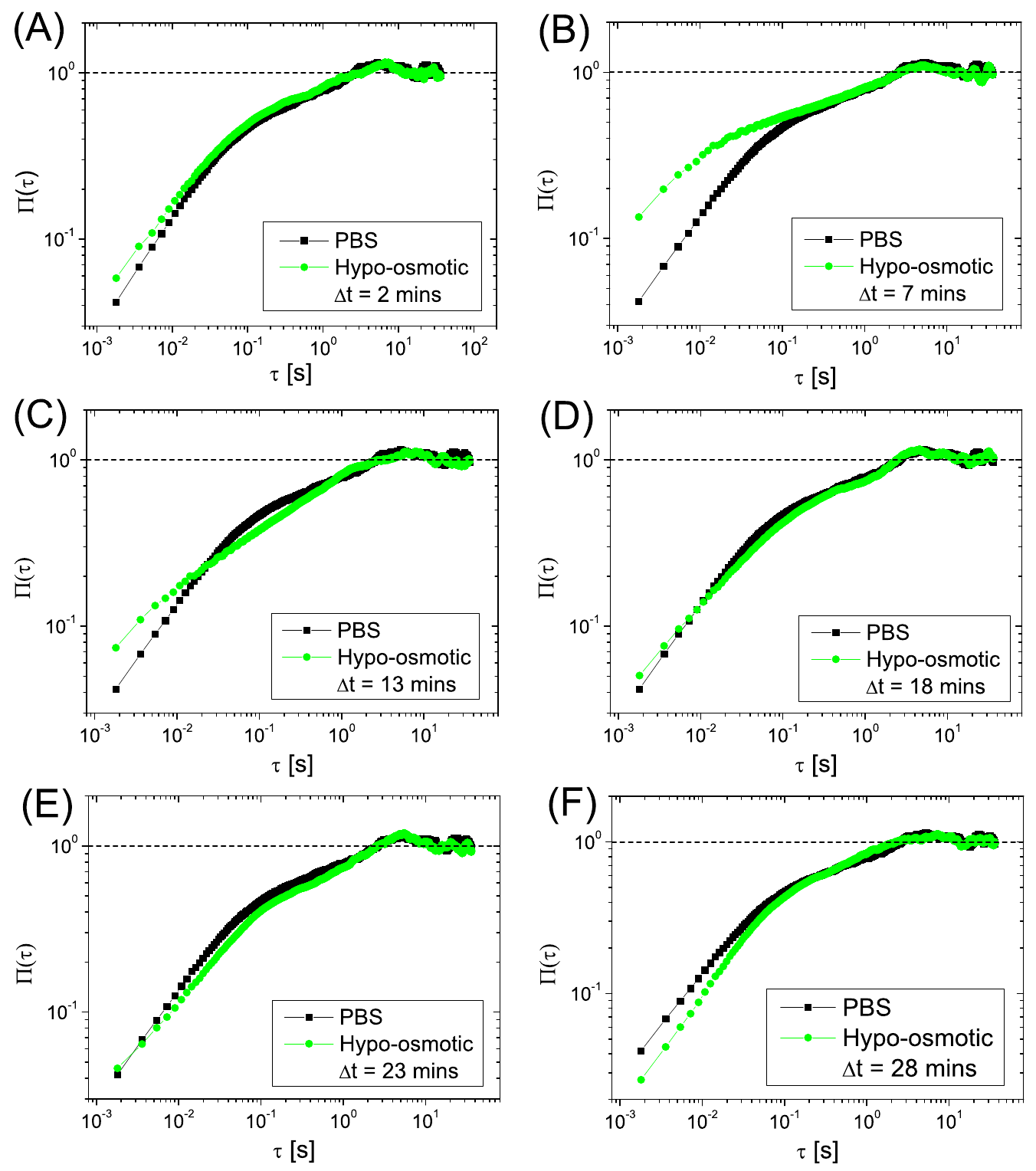}
\end{center}
\caption{
{\bf The normalised mean--square displacement \textit{vs.} lag--time of a 5 $\mu m$ diameter silica bead chemically bound to a Jurkat cell in iso--osmotic (PBS) solution (square symbols) and in hypo-osmotic solution (circle symbols) after the addition of 10\% v/v distilled water to the PBS buffer and measured at time intervals ($\Delta t$) of (A) 2 min, (B) 7 min, (C) 12 min, (D) 18 min, (E) 23 min, (F) 28 min, respectively.} At short time intervals, the $\Pi (\tau)$ has the potential to reveal the dynamics occurring at molecular level (as shown in Figure~\ref{Figure_8}); whereas, at large lag--times it provides information on the stiffness of the whole cell (as shown in Figure~\ref{Figure_4}).}
\label{Figure_6}
\end{figure}

In order to study the high frequency behaviour of cells, we have used Equation~\ref{G*c}, which directly relates the normalised mean-square displacement, $\Pi (\tau)$, of the bead to the cell complex modulus, $G^*_c (\omega)$. In particular, we have applied Equation~\ref{G*c} to the $\Pi (\tau)$ derived from measurements that were collected at fixed time intervals after the solution was made hypotonic. In Figure~\ref{Figure_6}, these are compared with the normalised MSD of the same cell in phosphate-buffered saline (PBS) solution, prior to hypotonic exposure. Note that in order to more easily discern visually the changes in $\Pi (\tau)$, caused by the hypotonicity, the normalised MSD for the cell in PBS is plotted in each of the panels.

\begin{figure}[!ht]
\begin{center}
\includegraphics[height=6.5in,width=4.5in]{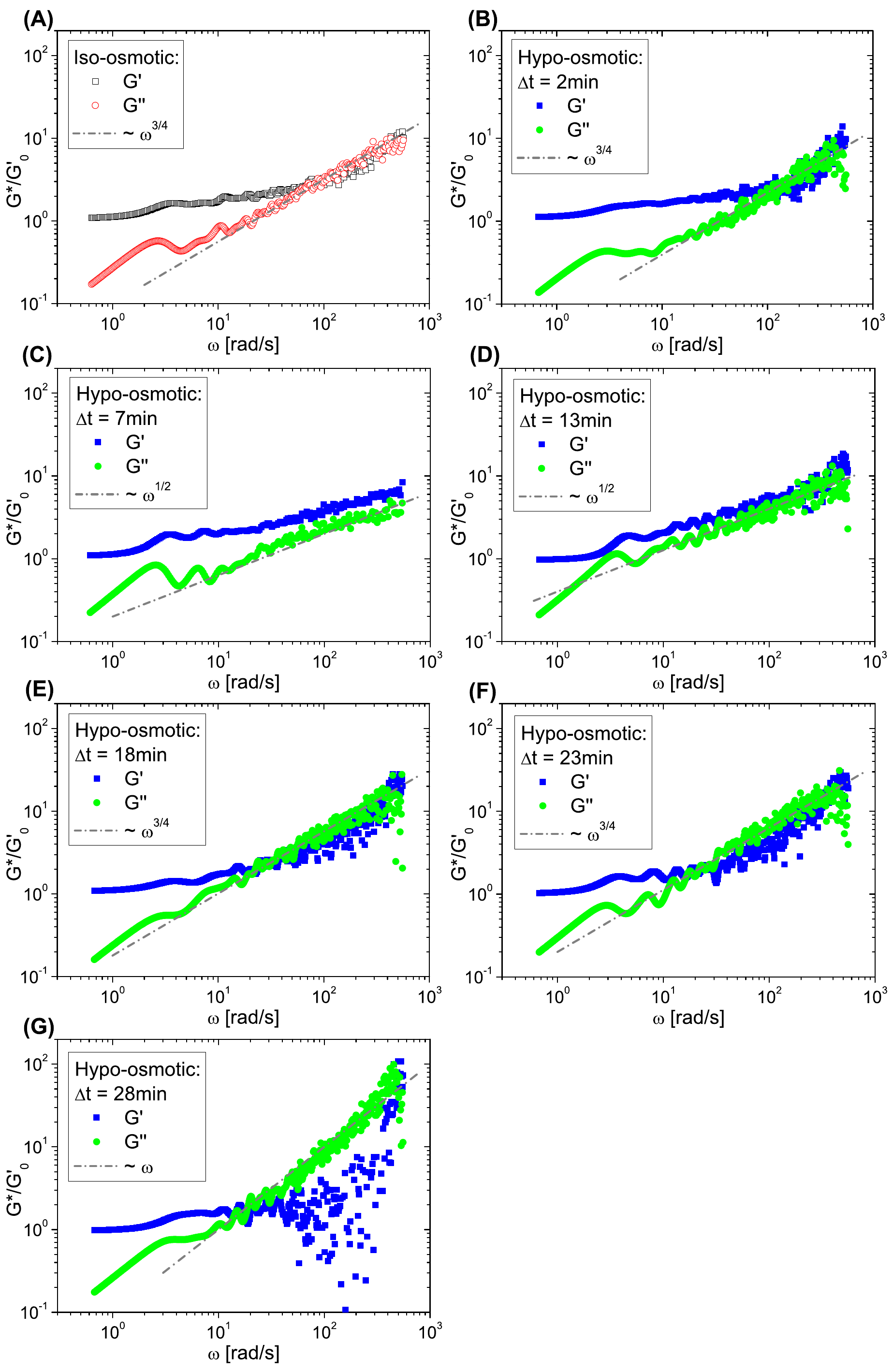}
\end{center}
\caption{
{\bf The real (storage, $G'_{c}(\omega)$) and the imaginary (loss, $G''_{c}(\omega)$) parts of the complex modulus ($G^{*}_{c}(\omega)$) scaled by $G'_{0}$ \textit{vs.} frequency of a Jurkat cell in iso-osmotic (PBS) solution (A) and in hypo-osmotic solution (B-G) after the addition of 10\% v/v distilled water to the PBS buffer and measured at time intervals ($\Delta t$) of (B) 2 min, (C) 7 min, (D) 12 min, (E) 18 min, (F) 23 min, (G) 28 min, respectively.} The moduli have been evaluated by using Equation~\ref{G*c} on the normalised mean-square displacement data shown in Figure~\ref{Figure_6}. The lines are guides for the gradients. Note that, the scatter in the data at high frequencies and the ripples in the low frequency portions of the curves in Figure~\ref{Figure_7} are due to the analytical method that we have adopted \cite{Evans:2009ay} for performing the Fourier transforms of the normalised MSDs. As explained in detail in Ref. \cite{Evans:2009dw}, this method works directly on the experimental data points (i.e. $\big\{\tau_k , \Pi_k \big\}$, where $k=1..N$) and therefore preserves genuine experimental noise.}
\label{Figure_7}
\end{figure}

The elastic ($G'_c (\omega)$) and viscous ($G''_c (\omega)$) components of the complex modulus, $G^*_c (\omega)$, derived from the measurements shown in Figure~\ref{Figure_6} are shown in Figure~\ref{Figure_7} (scaled by the plateau modulus $G'_c (0)$). It is clear that, in PBS (Figure~\ref{Figure_7}A), the high frequency behaviour of the cell's elastic and viscous moduli both show a frequency dependence of $\propto \omega^{3/4}$, which is characteristic of isotropic \textit{in vitro} reconstituted actin filament solutions \cite{FARGE:1993uf, Tassieri:2008kr}. 

As the solution is made hypo--osmotic, the cell cytoskeleton starts to reorganise and the frequency behaviour of the moduli drastically change as (sequentially) shown in Figure~\ref{Figure_8}. In particular, after $\thicksim$10min in the hypo--osmotic solution both moduli tend to assume a high frequency behaviour $\propto \omega^{1/2}$ (see Figures~\ref{Figure_7}C and~\ref{Figure_7}D). This change in the degree of elasticity may be explained by two alternative (and possibly concomitant) processes: ($i$) an increase in cytoskeletal tension as response to a stretching force caused by the cell swelling and ($ii$) the consequent formation of  ``\textit{dangling branches}" \cite{Rouse:ya, Ferry:1980jo, Rubinstein:2003fb} (Figure~\ref{Figure_5}) generated during the reorganisation process of the cell cytoskeleton (including those possibly obtained from the rupture of the cytoskeletal protein filaments because of stress overload). However, whilst the increase in cytoskeletal tension would explain both the initial increase of the cell stiffness and the change in the frequency scaling laws of the moduli (i.e. from $\propto \omega^{3/4}$ to $\propto \omega^{1/2}$), it would fail to explain the softening process occurring during the cell volume re--equilibration, which ends with a breakdown, at high frequency, of the cell's elastic modulus (Figures~\ref{Figure_7}G). This latter phenomenon could be simply justified by an \textit{excess} of ``\textit{dangling branches}". Indeed, from a rheological point of view, these would only contribute to the capability of the cell to dissipate energy but not to the cell elasticity, since they cannot bear stress and hence do not contribute to the cell's rigidity.

Finally, it is important to highlight that the results that have been obtained here using a video particle tracking method are in good agreement with those presented in the review written by Papakonstanti and Stournaras \cite{Papakonstanti:2007fi}, where a set of assessments of actin cytoskeleton dynamics and actin architecture in cell volume regulation are summarised. However, none of the techniques reviewed in that work \cite{Papakonstanti:2007fi} are able to provide quantitative information on the cell viscoelasticity, as is the case here.

\clearpage

\clearpage

\section*{Conclusions}
In summary, we have presented a straightforward and \textit{non--invasive} experimental procedure, coupled with a new analytical method to interpret the data, which leads to quantitative determination of the \textit{in vivo} viscoelastic properties of cells in the frequency domain. The method has the potential to monitor the internal dynamics and re--organisation of the actin cytoskeleton up to frequencies on the order of $kHz$, representing a valuable addition to studies that address cellular physiology and pharmacological response. Indeed, in this work we report, \textit{for the first time}, the high frequencies (up to $\thicksim$600Hz) changes of the Jurkat cells' viscoelastic spectrum from $\propto \omega^{3/4}$ to $\propto \omega^{1/2}$, as response to a change in osmolarity of the solution. The rheological interpretation of the results gives a direct insight of the cell cytoskeleton structure and its re--organisation. In the future, it is envisaged that these interpretations could be coupled with advanced molecular biology techniques to resolve the detailed interactions underlying these rheological changes and that faster dynamics could be studied by means of a quadrant photo--diode based tracking system

\section*{Acknowledgments}
We thank Mike Evans for helpful conversations. M.T. is Research Fellow of the Royal Academy of Engineering/EPSRC; research program title: ``\textit{Rheology at the Microscale: New Tools for Bio-analysis}". We are grateful to EPSRC and BBSRC for supporting this work through grants EP/F040857/1 and BB/C511572/1, respectively, and to the DTC in Proteomic and Cell Technologies (EPSRC) for funding RLW. 

\bibliographystyle{plos2009}
\bibliography{Jurkatcell}


\end{document}